\documentclass[
twocolumn, preprintnumbers,nofootinbib,amsmath,amssymb]{revtex4-1}

\usepackage{graphicx}
\usepackage{dcolumn}
\usepackage{bm}
\usepackage{soul}
\usepackage{xcolor}

\newcommand\ee{\end{equation}}
\newcommand\be{\begin{equation}}
\newcommand\eea{\end{eqnarray}}
\newcommand\bea{\begin{eqnarray}}

\newcommand{\de}{\partial}
\def\beq{\begin{equation}}
\def\eeq{\end{equation}}
\def\ep{\epsilon}

\newcommand{\bsb}{\boldsymbol}

\newcommand{\Or}{\mathcal{O}}

\newcommand{\z}{\zeta}

\newcommand{\q}{{\bsb q}}
\newcommand{\kb}{{\bsb k}}
\newcommand{\x}{{\bsb x}}
\newcommand{\s}{{\bsb s}}
\renewcommand{\z}{{\bsb z}}
\newcommand{\expect}[1]{\left\langle #1 \right\rangle}
\newcommand{\lb}{\ell_{\rm BAO}}
\newcommand{\comment}[1]{}

\renewcommand\[{\left[}

\usepackage[hidelinks]{hyperref} 
\comment{
\hypersetup{
colorlinks=true,
citecolor=DarkBlue,
linkcolor=DarkBlue,
urlcolor=DarkBlue,
}}

\begin{document}

\title{Equivalence Principle and the Baryon Acoustic Peak}

\author{Tobias Baldauf}
\author{Mehrdad Mirbabayi}
\author{Marko Simonovi\'c}
\author{Matias Zaldarriaga}
\affiliation{Institute for Advanced Study, Einstein Drive, Princeton, NJ 08540, USA}

\begin{abstract}
We study the dominant effect of a long wavelength density perturbation $\delta(\lambda_L)$ on short distance physics. In the non-relativistic limit, the result is a uniform acceleration, fixed by the equivalence principle, and typically has no effect on statistical averages due to translational invariance. This same reasoning has been formalized to obtain  a ``consistency condition'' on the cosmological correlation functions. In the presence of a feature, such as the acoustic peak at $\lb$, this naive expectation breaks down for $\lambda_L<\lb$. We calculate a universal piece of the three-point correlation function in this regime. The same effect is shown to underlie the spread of the acoustic peak, and is calculable to all orders in the long modes. This can be used to improve the result of perturbative calculations --- a technique known as ``infra-red resummation''--- and is explicitly applied to the one-loop calculation of power spectrum. Finally, the success of BAO reconstruction schemes is argued to be another empirical evidence for the validity of the results.

\end{abstract}
\maketitle

\section{Introduction}

Local experiments performed in a small laboratory cannot reveal the existence of the uniform gravitational field of a long-wavelength matter density perturbation, e.g. $\delta_L(\x,t) = \delta_\q(t) \cos(\q\cdot \x)$. By the equivalence principle, the laboratory and all of its belongings fall with a uniform acceleration $-\nabla \Phi_L(\x_{\rm lab},t)$, where $\Phi_L(\x,t)=-4\pi G a^2 \bar\rho(t)\delta_L(\x,t)/q^2$, and $\bar \rho(t)$ is the mean matter density of the Universe. However, two distant laboratories with separation larger than $1/q$ experience different accelerations. A distant observer sees a clear correlation between the relative motion of the two and the underlying density perturbation.

The motion in the field of a long-wavelength mode is easiest to find from the fact that everything falls in the same way as a dark matter particle does. Possible deviations are suppressed by additional derivatives of the long mode. For dark matter, the linearized continuity equation implies $\bsb v \simeq -\frac{\nabla}{\nabla^2}\dot\delta$. The total displacement since $t=0$ is then
\be\label{Dx}
\Delta \x=\delta_\q(t) \sin(\q\cdot \x) \, \q/q^2.
\ee

The small laboratories of the cosmologist, like stars and galaxies, are observed at a single point in their lifespan. Hence, the relative motion of any given pair is impossible to determine. What is possible is to see how the distribution of pairs is correlated with $\delta_L$. For pairs of any objects, say galaxies, equation \eqref{Dx} implies
\be
\label{xi}
\begin{split}
\expect{\delta_g(\frac{\x}{2},t)\delta_g(-\frac{\x}{2},t)}_{\delta_L}&\simeq \xi_g(\x,t)\\[8pt]
+2\delta_\q(t) &\sin\left(\frac{\q\cdot\x}{2}\right) \frac{\q}{q^2}\cdot \nabla \xi_g(\x,t),
\end{split}
\ee
where $\xi_g(\x,t)$ is an average 2-point correlation function. Not surprisingly, the distribution of pairs with separation much less than the long wavelength, $\q\cdot\x\ll 1$, is hardly effected by the long mode. The second line would in this case correspond to the effect of living in an over (under) dense Universe. An effect of order $\delta_L x |\nabla \xi_g|$, which for an approximately scale invariant spectrum, $|\nabla \xi_g(\x,t)| \sim \xi_g(\x,t)/x$, is comparable to dynamical contributions of order $\delta_L \xi_g$, which are neglected anyway on the right-hand side (r.h.s.). However, even if $\q\cdot\x\gg 1$, when we do expect the long-wavelength mode to induce a large relative motion, the second line of \eqref{xi} is often negligibly small. Scale invariance now implies that it is of order $\delta_L \xi_g /q x$. 

The relative motion is noticeable only if the distribution of pairs has a feature such that the derivative in the second line of \eqref{xi} becomes large. One such feature does exist in the Universe at the baryon acoustic oscillation (BAO) peak. For $x\sim \lb$, 
\be
|\nabla \xi_g|\sim\frac{1}{\sigma}\xi_g \gg \frac{1}{\lb}\xi_g,\\[10pt]
\ee
where $\sigma$ is the width of the peak. At this separation, the effect of the long mode on the distribution of pairs is of order $\delta_L \lb\xi_g/\sigma$ for $q\ll \lb^{-1}$, and $\delta_L \xi_g /q \sigma$ for  $\lb^{-1}\ll q\ll\sigma^{-1}$, which are both dominant compared to the $\Or(\delta_L\xi_g)$ dynamical effects. In what follows, we explore the implications of this simple observation for the shape of the correlation functions around the BAO scale, and its connection to broadening of the peak.\footnote{The initial time for this problem can be taken long after the recombination, when the acoustic peak is already in place, but the modes of interest, including those actually forming the peak, are still linear and Gaussian.}

\section{Correlation with the long mode}

\noindent
{\bf Real space.---}
An approximate three-point correlation function can be obtained in this regime by correlating \eqref{xi} with $\delta(\q,t)$ to get
\be
\label{eq:real}
\begin{split}
\expect{\delta(\q,t) \delta_g(\frac{\x}{2},t)\delta_g(-\frac{\x}{2},t)}&\\[8pt]
\simeq 2 P_{\rm lin}(q,t) \sin &\left(\frac{\q\cdot\x}{2}\right) \frac{\q}{q^2}\cdot \nabla \xi_g(\x,t),
\end{split}
\ee
where $P_{\rm lin}(q,t)$ is the linear matter power spectrum.\footnote{We use finite volume Fourier transformation where the cosine mode is related to Fourier modes according to $\delta_\q(t)=[\delta(\q,t)+\delta(-\q,t)]/V$. However, for convenience, the discrete momentum sums are approximated by integrals $V\int {\rm d}^3\q/(2\pi)^3$. Thus,
\be
\expect{\delta(\q,t)\delta(\q',t)}=P(q) (2\pi)^3\delta^3(\q+\q'),\nonumber
\ee
with $\delta^3({\bsb 0})\equiv V/(2\pi)^3$. In what follows, momentum conservation is always explicitly imposed on momentum-space correlation functions, but the factor $(2\pi)^3\delta^3(\sum \q_i)$ is dropped. Note also that the sine modes do not contribute to the relative displacement of pairs located at $\pm \frac{\x}{2}$.}

\begin{figure}[!t]
\begin{center}
\includegraphics[width=0.48 \textwidth]{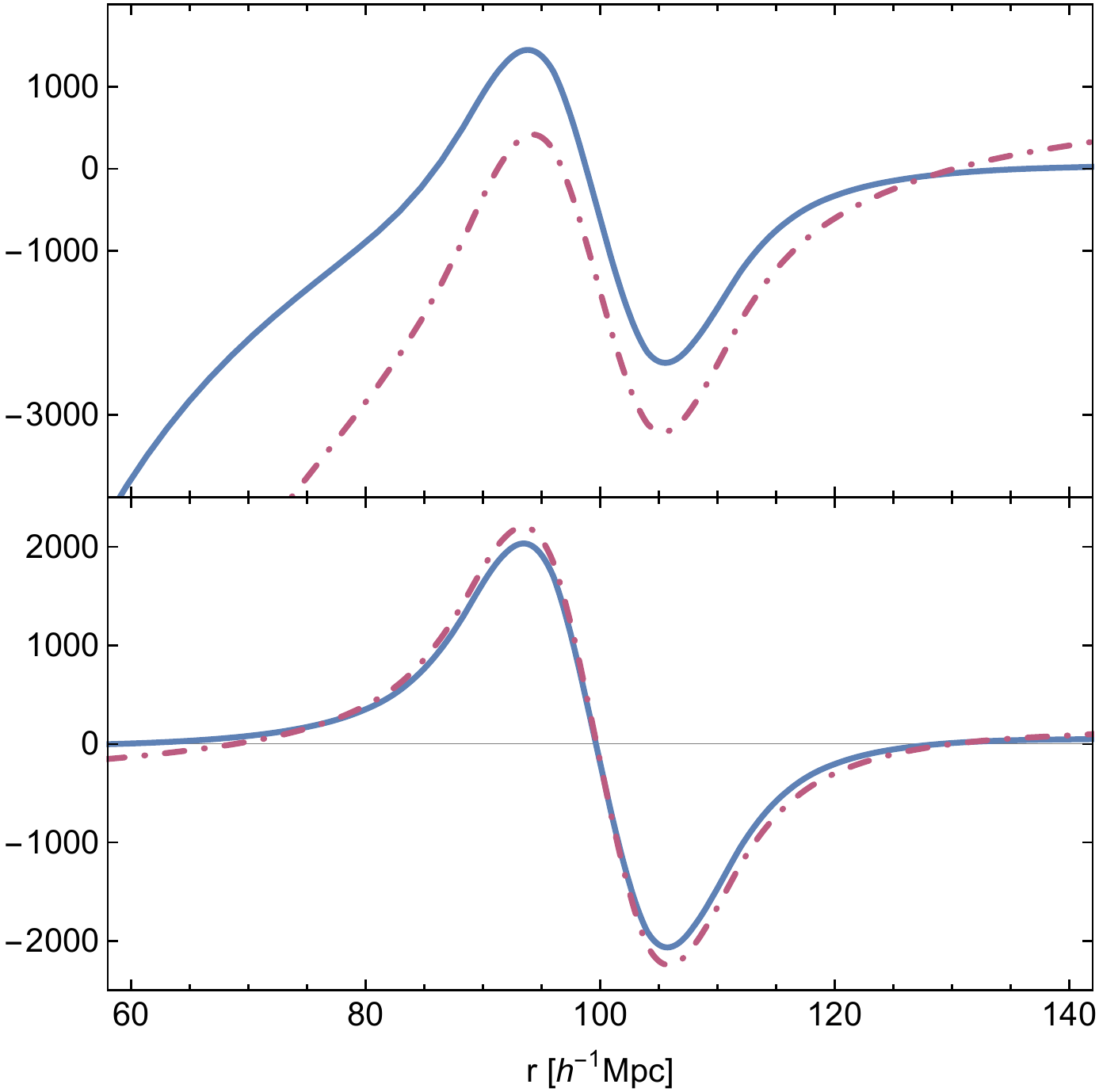}
\end{center}
\caption{\small {\em Upper panel: The mixed real-momentum space three-point function of equation \eqref{eq:real} (solid line) and the perturbation theory result (dot-dashed line) as a function of $r$. Both curves are obtained for $q=0.03\;h{\rm Mpc}^{-1}$, and are normalized by $P_{\rm lin}(k_{\rm eq})\xi(2\pi/k_{\rm eq})$. Lower panel: The comparison between the two results when the background (calculated from the featureless power spectrum) is subtracted.}}
\label{fig:mixed}
\end{figure}

Given that terms of order $P_{\rm lin}\xi_g$ have been neglected from the r.h.s., one must ask how accurate is the above approximation. Realistically, $\xi_g$ contains also a smooth background. Hence, the approximation is valid so long as 
\be
\sin \left(\frac{\q\cdot\hat\x}{2}\lb\right) \frac{\q\cdot\hat\x}{\sigma q^2}\xi^w_g\gg \xi_g,
\ee
where $\xi_g^w$ is the ``wiggle'' component. In our Universe, $\xi_g^{w}/\xi_g\approx 0.8$, the peak location $\lb\approx 100 h^{-1}$Mpc, and the width $\sigma \approx 10 h^{-1}$Mpc. Therefore, the corrections are of order 10-20\%, but become larger at the nodes of the sine and as $q\to 2\pi\sigma^{-1}$. 

Note, however, that what is more essential in the above derivation is the breakdown of scale invariance characterized by $\sigma/\lb\ll 1$, rather than the actual size of the feature. Even for small $\xi^w_g/\xi_g$, the contribution \eqref{eq:real}, with $\xi_g$ replaced by $\xi_g^w$, is distinct though perhaps subleading. 

We tested the above expectations by taking $\delta_g$ to be the matter contrast itself, and using the tree-level expression for the bispectrum in perturbation theory---fig.~\ref{fig:mixed}.\footnote{The tree-level bispectrum is given by
\be
B(\kb_1,\kb_2,\kb_3)= 2[F_2(\kb_1,\kb_2)P(k_1)P(k_2)+\text{2 permutations}],
\ee
where $F_n$ are the usual Standard Perturbation Theory (SPT) kernels \cite{spt}:
\be
F_2(\kb_1,\kb_2)=\frac{5}{7}+\frac{2}{7}\frac{(\kb_1\cdot\kb_2)^2}{k_1^2k_2^2}+\frac{1}{2}(\kb_1\cdot\kb_2)
\left(\frac{1}{k_1^2}+\frac{1}{k_2^2}\right).
\ee
For simplicity, the plots are made using the BBKS power spectrum \cite{Bardeen:1985tr} modified to account for BAO wiggles
\be\label{P}
P(k)=P_{\rm BBKS}(k)(1+T^w(k/k_{\rm eq.}))\;,
\ee
where $k_{\rm eq}=0.01 h{\rm Mpc}^{-1}$ is the equality scale, and the transfer function $T^w(x)$ is given by
\be
T^w(x)= a\sin(f x)W(x,x_{\rm max})(1-W(x,x_{\rm min})) \;,\nonumber
\ee
where $W(x,x_0)=\exp(-x^2/x_0^2)$. The parameters are chosen to reproduce the observed BAO peak: $a=0.05$, $f=1$, $x_{\rm max}=30$, and $x_{\rm min}=3$. 
} As seen, subtracting the smooth contribution of the background results in a much better agreement.

\vspace{0.3cm}
\noindent
{\bf Squeezed limit bispectrum.---}
Taking the Fourier transform of \eqref{eq:real} with respect to $\x$, we obtain the squeezed limit ($q\ll k$) momentum space bispectrum:
\be\label{eq:momentum}
\begin{split}
\expect{\delta(\q,t) \delta_g(\kb_-,t)\delta_g(-\kb_+,t)}&\\[8pt]
\simeq \frac{\q\cdot \kb}{q^2}\ P_{\rm lin}(q,t)&[P_g(k_-,t)- P_g(k_+,t)],
\end{split}
\ee
where $\kb_\pm\equiv \kb\pm \q/2$. The above derivation can be generalized to the case where the fields have different time arguments. The result, often called the squeezed limit consistency condition (see e.g. \cite{Kehagias:2013yd, Peloso:2013zw, Creminelli:2013mca}), reads
\be
\label{eq:consistency}
\begin{split}
\expect{\delta(\q,t) \delta_g(\kb_-,t_1)\delta_g(-\kb_+,t_2)}& \simeq \frac{\q\cdot \kb}{q^2}\\[8pt]
P_{\rm lin}(q,t)\Big[\frac{D(t_1)}{D(t)} P_g(k_-,t_1)&- \frac{D(t_2)}{D(t)} P_g(k_+,t_2) \Big] \;,
\end{split}
\ee
where $D(t)$ is the linear growth factor. The $k_\pm$ in the arguments of $P_g$ are normally approximated by $k$, which is valid in the $q\to 0$ limit: the difference
\be\label{d/dk}
P_g(k_\pm,t)-P_g(k,t)=\pm \frac{1}{2}\q\cdot\nabla_\kb P_g(k,t)
\ee
results in an $\Or(q^0)$ term in \eqref{eq:consistency} that is comparable to other dynamical effects of the long mode. This has led to the conclusion that the $1/q$ contribution to the squeezed limit bispectrum vanishes at equal times. (The $1/q$ piece survives in unequal time correlations. Measuring unequal time correlations is equivalent to watching the galaxies as they fall in the long wavelength gravitational field. Unfortunately, this is practically impossible.) 

However, the above reasoning does not necessarily hold when considering squeezed triangles with small but finite $q$. In the presence of the acoustic feature, $P_g(k,t)$ has an oscillatory component with period $2\pi \lb^{-1}$, which can be smaller than $q$. In this regime, the approximation \eqref{d/dk} is invalid and the difference is proportional to the power $P_g^w(k,t)$ in the acoustic peak -- the Fourier transform of $\xi_g$ after the subtraction of a smooth background. The r.h.s. of \eqref{eq:momentum} now reads
\be\label{bi}
2 P_{\rm lin}(q,t) \sin\left(\frac{\q\cdot\hat\kb}{2}\lb\right) \frac{\q\cdot \nabla_{\hat\kb}}{\lb q^2} P_g^{w}(k,t),
\ee
where $\nabla_{\hat\kb}\equiv k \nabla_\kb$. To derive this expression, we have used the fact that the Fourier transform of a sharp feature is generically a fast oscillating piece times a smooth envelope; equation \eqref{P} is an example. For $q\lb\gg 1$ the result is enhanced by a factor of $k/q$.\footnote{In the case of higher point correlation functions \eqref{eq:momentum} generalizes to
\be
\begin{split}
&\expect{\delta(\q,t) \delta_g(\kb_1,t)\cdots\delta_g(\kb_n,t)}\simeq 
P_{\rm lin}(q,t) \\
&~~~~~~~~~~~~~\sum_i \frac{\q\cdot \kb_i}{q^2}\expect{\delta_g(\kb_1,t)\cdots\delta_g(|\kb+\q|,t)\cdots\delta_g(\kb_n,t)},\nonumber
\end{split}
\ee
which again scales as $1/q$ for $q>2\pi\lb^{-1}$.}

\begin{figure}[!t]
\begin{center}
\includegraphics[width=0.48 \textwidth]{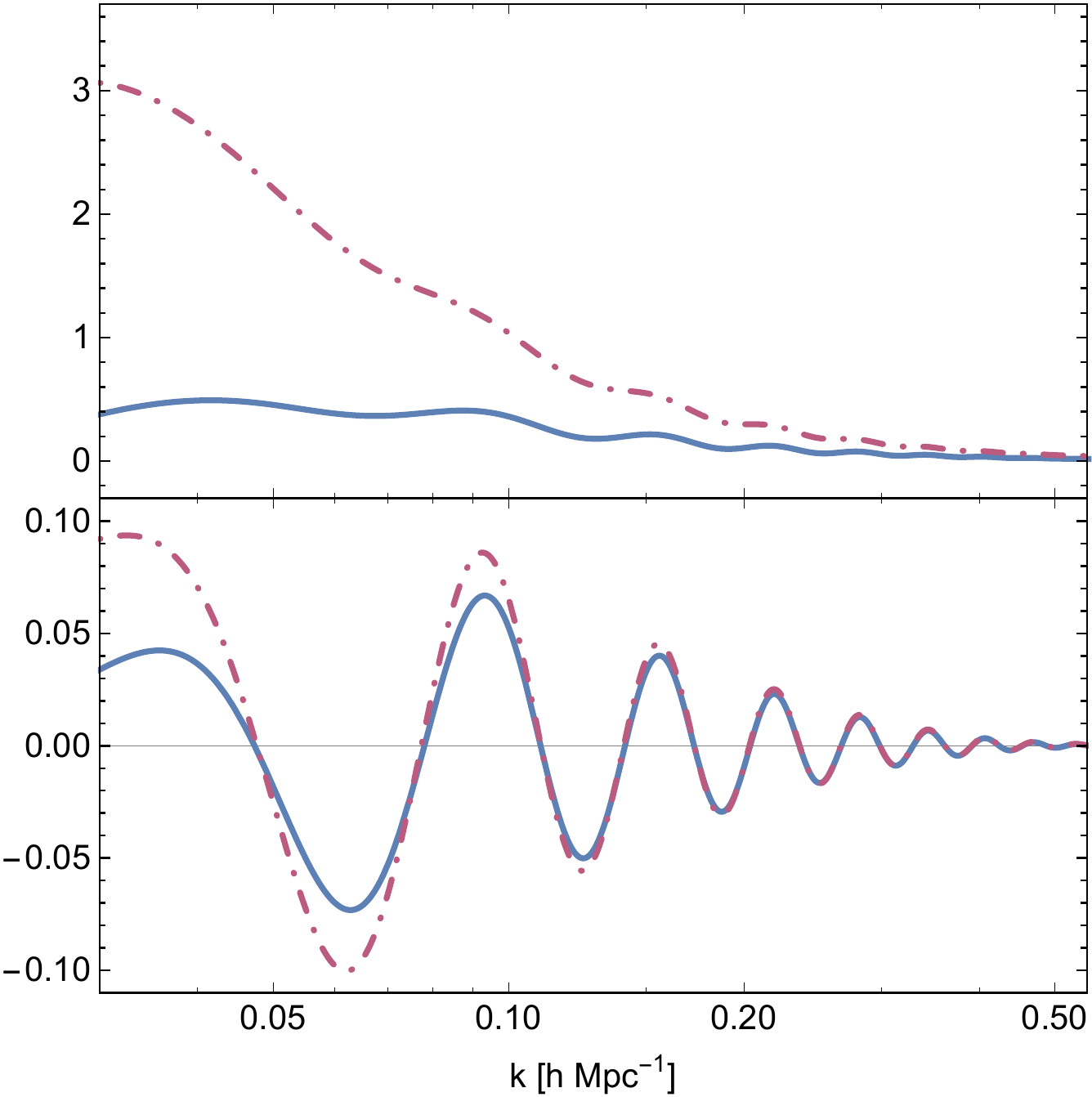}
\end{center}
\caption{\small {\em Upper panel: The bispectrum calculated using equation \eqref{bi} (solid line) and the tree-level perturbation theory result (dot-dashed line) as a function of $k$, for $q=0.03\;h{\rm Mpc}^{-1}$. Both curves are normalized by $P_{\rm lin}^2(k_{\rm eq})$. Lower panel: The same as above with the smooth background subtracted.}}
\label{fig:momentum}
\end{figure}

Nevertheless, compared to other dynamical contributions, expression \eqref{bi} is suppressed by $P_g^{w}(k)/P_g(k)$. In the case of the initial matter power spectrum this ratio has support for $k\lb < 100$ and reaches a maximum of approximately $0.05$ at $k\lb\sim 10$. The overall result turns out to be a subdominant component of the full momentum space bispectrum, essentially because most of the power at high $k$ comes from short distance correlations $\xi_g(x\sim 2\pi/k)$ rather than the acoustic feature. A comparison with the tree-level matter bispectrum is shown in fig.~\ref{fig:momentum}. As seen, once the smooth background is subtracted, what remains is well approximated, in the squeezed limit, by the universal result \eqref{bi}.

\section{BAO spread and reconstruction}

Intuitively, the above result describes how galaxy pairs, which are more likely to be found at distance $\lb$, are moved to larger or smaller separations in the presence of a mode of wavelength longer than $\sigma$. When averaged over the long modes, these motions lead to the well-known spread of the acoustic peak, as will be discussed in the rest of the paper.

For this purpose, it is necessary to keep higher order terms in the expansion \eqref{xi}. At second order in relative displacement, now caused by the modes $\q_1$ and $\q_2$, the r.h.s. reads
\be\label{2nd}
2\delta_{\q_1}\delta_{\q_2} \sin\left(\frac{\q_1\cdot\x}{2}\right)
\sin\left(\frac{\q_2\cdot\x}{2}\right) \frac{q_1^iq_2^j}{q_1^2q_2^2}\: \de_i\de_j \xi_g(\x,t).
\ee
As before, this is the leading effect of the long mode if $x\approx\lb$, and $\xi_g$ is the correlation function in the absence of the $q$ modes. By correlating \eqref{2nd} with two long modes one can obtain the double-squeezed four-point correlation function. Alternatively, averaging over the long modes with $q<\Lambda\ll 2\pi\sigma^{-1}$, gives the first correction to the observed two-point correlation around the peak:
\be\label{1loop}
\tilde \xi_g(r,t)\approx \xi_{g,L}(r,t)+\xi_{g,S}(r,t)+\Sigma_\Lambda^2\xi_{g,S}''(r,t), 
\ee
where $r\equiv |\x|$, prime denotes $\de/\de r$, and terms suppressed by $\sigma/\lb$ are neglected. $\xi_{g,L}(\x,t)$ -- the direct contribution of the long-modes to the correlation function -- and $\xi_{g,S}(\x,t)$ -- that of the short modes in the absence of the long modes-- are assumed to be isotropic. Note that while $\xi_{g,S}$ contains the full short scale nonlinearities, only the leading effect of the long modes on the short modes has been kept in \eqref{1loop}. For each $q$ mode, this scales as $P_{\rm lin}(q)(\lb/\sigma)^2$ for $q\ll \lb^{-1}$, and $P_{\rm lin}(q)/(q\sigma)^2$ for $q>\lb$. The corrections are suppressed by one or more powers of $\sigma/\lb$ and $q\sigma$, respectively. Hence, due to the bulk motions, $\tilde \xi_g$ has a broader peak with $\Sigma_\Lambda^2$ given by
\be\label{Sigma}
\Sigma_\Lambda^2 \approx \frac{1}{6\pi^2}\int_0^{\Lambda}\!\!\! {\rm d} q \: P_{\rm lin}(q)[1-j_0(q\lb)+2j_2(q\lb)],
\ee
where $j_n$ is the $n^{th}$ order spherical Bessel function.

It is easy to perturbatively confirm the above result when $\xi_g$ is taken to be the dark matter correlation: The leading contribution of the long wavelength modes to the one-loop power spectrum of the peak reads\footnote{The full one-loop power spectrum is given by
\be\label{1loopspt0}
\int \frac{{\rm d}^3\q}{(2\pi)^3} [6F_3(\q,-\q,\kb)P_{\rm lin}(k)+2F^2_2(\q,\kb-\q)P_{\rm lin}(|\kb-\q|) ]P_{\rm lin}(q)\;.
\ee
For $q\ll k$ it reduces to \eqref{1loopspt}. Incidentally, this coincides with 
\be
\frac{1}{2}\int_{q\ll k}\!\!\frac{{\rm d}^3\q}{(2\pi)^3}\: 
P_{\rm lin}^{-1}(q)\expect{\delta_{\q}\delta_{-\q}\delta_\kb\delta_{-\kb}},\nonumber
\ee
as expected from the remark after \eqref{2nd}.}
\be\label{1loopspt}
\begin{split}
P_{\rm 1-loop}^w(k>\Lambda)=&\frac12\int^\Lambda\!\! \frac{{\rm d}^3\q}{(2\pi)^3} \:\frac{(\q\cdot\kb)^2}{q^4}P_{\rm lin}(q)\\[8pt]
[P^w_{\rm lin}(|\kb+\q|)&+P^w_{\rm lin}(|\kb-\q|)-2P^w_{\rm lin}(k)]\;.
\end{split}
\ee
For $q\ll k$ the expression in the square brackets simplifies to $-4 P^w_{\rm lin}(k)\sin^2 (\q\cdot\hat\kb\lb/2)$, giving
\be\label{Sigep}
P_{\rm 1-loop}^w(k>\Lambda)=\Sigma_\Lambda^2 k^2 P_{\rm lin}^w(k),
\ee
and taking the Fourier transform with respect to $\kb$ reproduces \eqref{1loop}.

Note that for any $k$, our approximation is valid for all $q\ll k$ while the above expressions are based on a rigid separation of scales above and below $\Lambda$. Of course, in reality $P_g^w(k)$ has support in a large range of momenta, roughly $(0.05-1)\; h{\rm Mpc}^{-1}$. Even if a $q$-mode falls in this range, it is still true that its leading effect on higher $k$ modes is the mere bulk motion. Therefore, it contributes to the peak power through $\xi_{g,L}$, and at the same time, broadens it by dispersing the shorter modes. A better estimate of the width can be obtained by including for each $k$ the broadening effect of all smaller $q$ modes, i.e. by taking $\Lambda$ to increase with $k$. Below, we will implement this idea by taking $\Lambda =\ep k$, with $\ep\ll 1$.

\begin{figure}[!t]
\centering
\includegraphics[width=0.48 \textwidth]{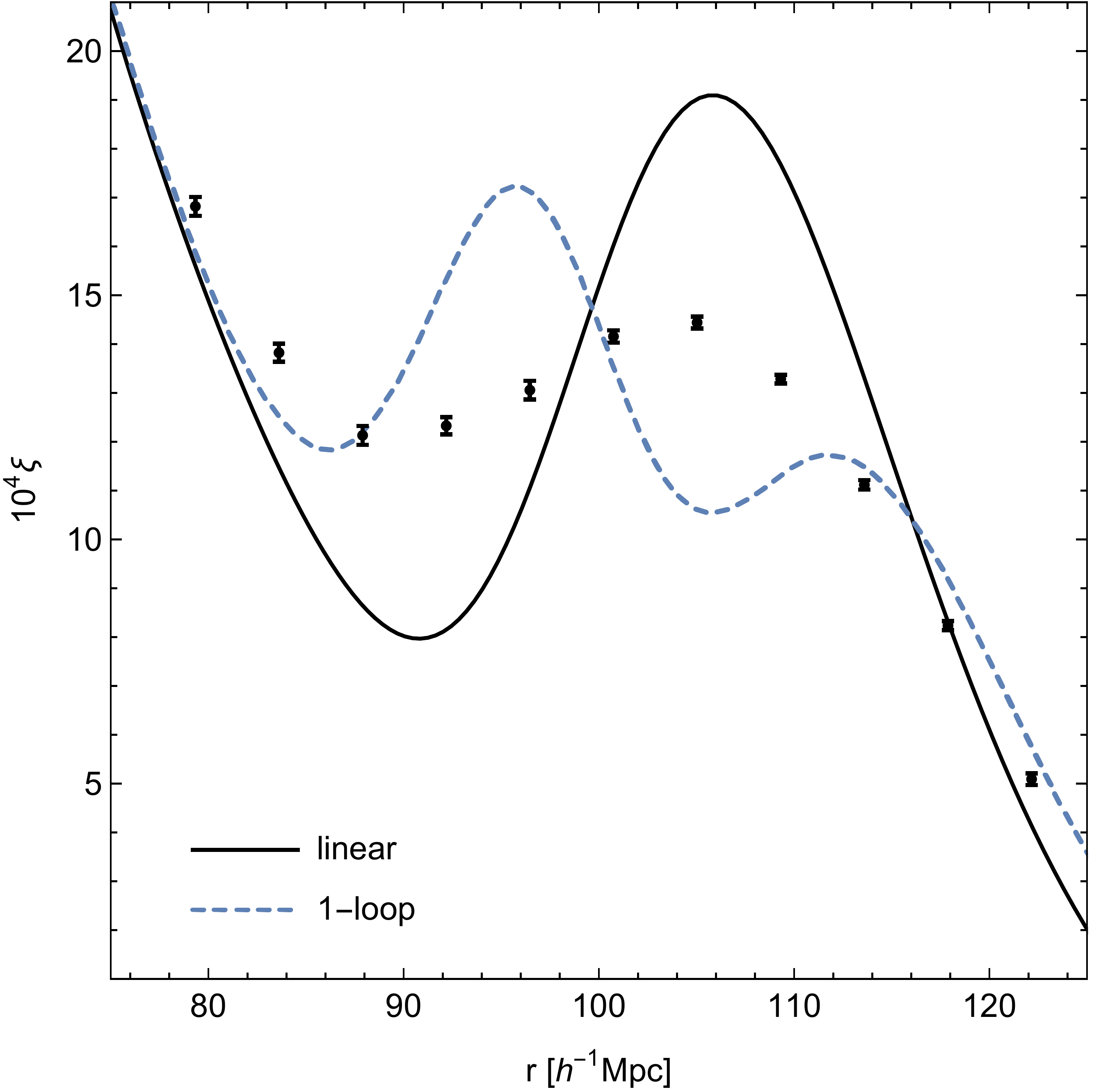}
\caption{\small {\em The acoustic peak in the matter correlation function in linear theory (solid), 1-loop perturbation theory (dashed), and simulation. }}
\label{1Lspread}
\end{figure}

Taking $\ep =1/2$, the above expression \eqref{Sigep} predicts an effective broadening of $\Sigma_{\ep k_*}\approx 5.5 h^{-1}\rm {Mpc}$, where $k_*$ is defined by $\Sigma_{\ep k_*}k_*=1$. This turns out to be a sizable fraction of the actual width of the observed matter correlation function. We compare the theoretical prediction with the result of an $N$-body simulation\footnote{We are measuring power spectra and correlation functions in a suite of 16 dark matter only simulations, each of which captures the evolution of $1024^3$ particles in a box of $1500^3\; h^{-3}\text{Mpc}^3$. The matter density parameter is $\Omega_\text{m}=0.272$, the tilt $n_\text{s}=0.967$ and the normalization $\sigma_8=0.81$. The leading cosmic variance has been divided out, such that the error bars reflect the sub-leading cosmic variance.} in fig.~\ref{1Lspread}. It is seen that the perturbative treatment has completely deformed the shape of the peak. A more accurate description should, therefore, treat the relative motions non-perturbatively.

\vspace{0.3cm}
\noindent
{\bf Infra-red resummation.---} We can obtain a formula which is valid to all orders in the relative displacement $\delta_\q/q$, by rewriting \eqref{xi} as (see e.g. \cite{Creminelli:2013poa})
\be
\label{xi2}
\begin{split}
\expect{\delta_g(\frac{\x}{2},t)\delta_g(-\frac{\x}{2},t)}_{\delta_L}&
\simeq \int \frac{{\rm d}^3\kb}{(2\pi)^3}\: e^{i\kb\cdot\x}\\[8pt]
\exp\Big[2i\delta_\q(t) \sin\left(\frac{\q\cdot\x}{2}\right)& \frac{\q\cdot\kb}{q^2}\Big]
\expect{\delta_g(\kb,t)\delta_g(-\kb,t)}.
\end{split}
\ee
As before, this is only relevant in the presence of a feature. Taking the expectation value over the realizations of the $q$ modes, approximating them, as we did so far, as being Gaussian, and using $\expect{\exp (i\varphi)}=\exp(-\expect{\varphi^2}/2)$ for Gaussian variables, we obtain our final expression for the dressed two-point correlation function around $r\approx \lb$
\be\label{dressed}
\tilde\xi_g(\x)\simeq \int \frac{{\rm d}^3\kb}{(2\pi)^3} \:e^{i\kb\cdot\x}e^{-\Sigma_{\ep k}^2 k^2}
\expect{\delta_g(\kb,t)\delta_g(-\kb,t)}_\ep.
\ee
To write the exponent in the above form, we have used the fact that $\nabla^2\approx \de_r^2$ [and therefore $k^2\approx (\hat\x\cdot\kb)^2$] up to corrections of order $\sigma/\lb$. In principle, the exponential factor should only multiply the peak power $P_g^w(k)$, though in practice the smooth background at $r\approx \lb$ is insensitive to the presence of this factor since $\Sigma\ll \lb$. The subscript $\ep$ on the momentum space expectation value on the r.h.s. indicates that it should be evaluated in the absence of modes with momentum $q$ smaller than $\ep k$, though it contains all short scale nonlinearities. Within a perturbative framework, it is possible to include dynamical effects of the long modes, as well as their non-Gaussianity by writing more complicated expressions (see below).

 To get an idea of how well \eqref{dressed} performs, we set $\delta_g=\delta$ and approximate the exclusive expectation value in the integral first by the linear matter power spectrum, and then by the 1-loop perturbation theory result. The first approximation underestimates the broadening by neglecting short scale nonlinearities and therefore predicts a slightly sharper peak. 

\begin{figure}[!t]
\centering
\includegraphics[width=0.48 \textwidth]{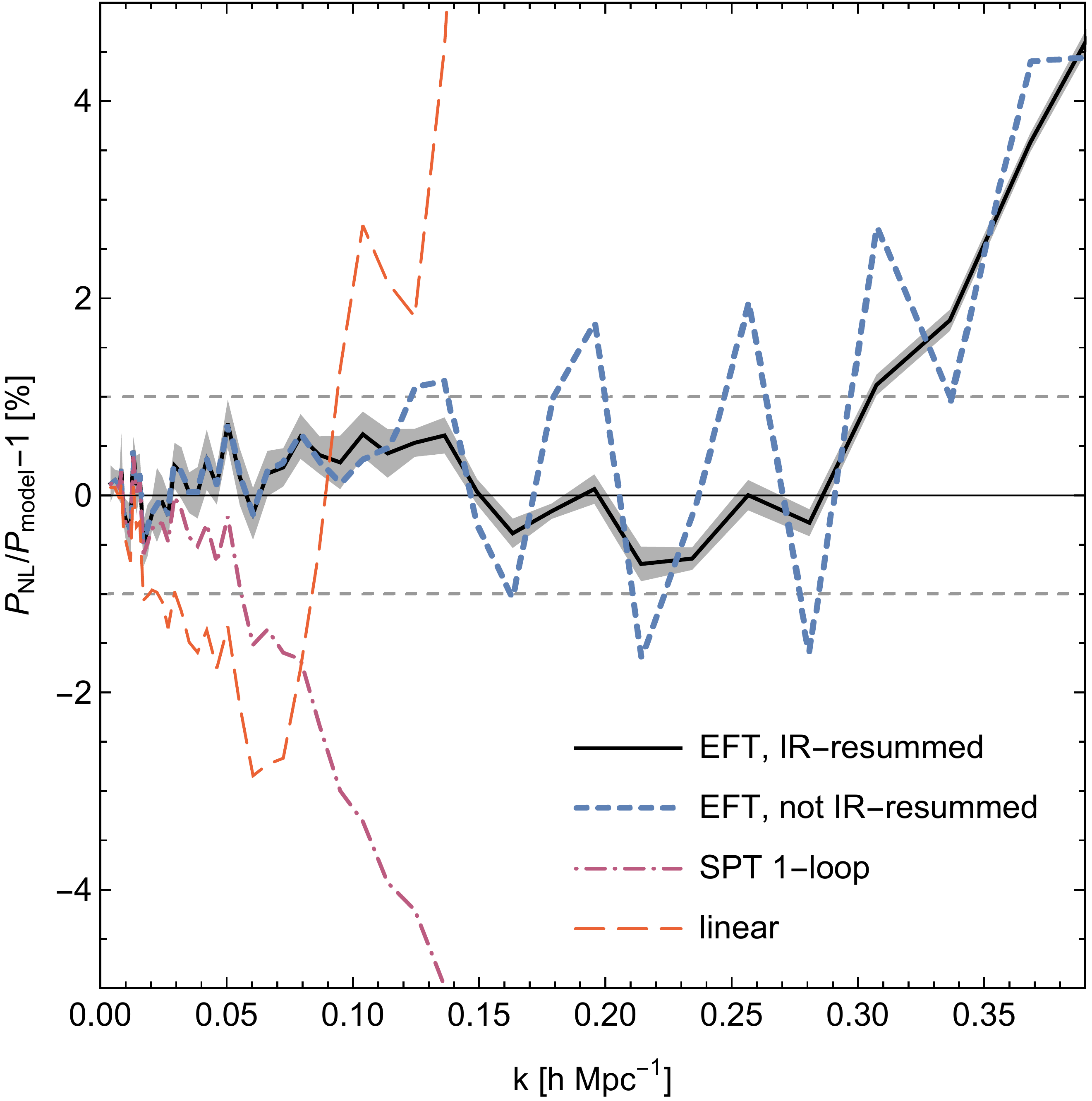}
\caption{\small {\em The ratio of various theoretical approximations to the power spectrum to the simulation result. Solid: IR-resummed \eqref{1loop1}, short-dashed: 1-parameter 1-loop EFT \eqref{1loop2}, dot-dashed: 0-parameter 1-loop EFT \eqref{1loop2} with $R=0$, and long-dashed: linear. The gray shaded region on the IR-resummed EFT curve gives the statistical error.}}
\label{fig:power}
\end{figure}

Let us discuss the 1-loop approximation in more details to see how \eqref{dressed} can be used to improve perturbative results. Two points have to be kept in mind: (i) The broadening is only relevant for the acoustic peak, hence the exponential broadening in \eqref{dressed} multiplies $P_\ep^w(k)$. (ii) Replacing $P^w_\ep(k)$ with the 1-loop power spectrum double-counts the effect of the long modes since the 1-loop result already contains $\Sigma^2_{\ep k} k^2 P_{\rm lin}^w(k)$ [c.f. \eqref{Sigep}]. Hence in this context the infra-red resummed version of the 1-loop power spectrum presented in \cite{Senatore:2014via} can be simplified and written as:
\be
\label{1loop1}
\begin{split}
\tilde P(k)&=P_{\rm lin}^{nw}(k)+P_{\rm 1-loop}^{nw}(k)\\[8pt]
+e^{-\Sigma^2_{\epsilon k}k^2}&(1+\Sigma^2_{\epsilon k}k^2)P^w_{\rm lin}(k)
+e^{-\Sigma^2_{\epsilon k}k^2} P^w_{\rm 1-loop}(k),
\end{split}
\ee
where the first line contains just the smooth part of the power spectrum.\footnote{In practice, $P_{\rm 1-loop}^{nw}$ can be obtained by substituting $P_{\rm lin}(k)$ with its no-wiggle part in the loop integrals \eqref{1loopspt0} since $P^w_{\rm lin}/P^{nw}_{\rm lin}\ll 1$.} When considering loop integrals with large internal momenta, one should allow for the possibility of higher derivative corrections to the dark matter equations of motion in an Effective Field Theory (EFT) framework \cite{Carrasco:2012cv}. These corrections compensate for the error made in treating the short-scale modes as a perfect fluid. Therefore, the EFT 1-loop power spectrum differs from \eqref{1loopspt0} by one such correction:
\be\label{1loop2}
P_{\rm 1-loop}(k)= P_{13}(k)+P_{22}(k)-2  R^2 k^2 P_{\rm lin}(k),
\ee
where $R$ (also known as speed of sound) is chosen to be $1.8\;h^{-2}{\rm Mpc}^2$ in order to obtain 1\% agreement with the simulation results up to $k_{\rm max} = 0.3 h\rm {Mpc}^{-1}$ (see fig.~\ref{fig:power}). This choice is a rough estimate of $R$, made in order to illustrate how the resummation improves matching the BAO oscillations for $k>0.1 h\rm {Mpc}^{-1}$. The exact value of $R$ is irrelevant for the shape of the acoustic peak.

The above resummation formula \eqref{1loop1} can be straightforwardly extended to any order in perturbation theory and to higher order statistics such as the bispectrum or trispectrum. Note that in this approximation the leading dynamical effect of the long modes on short modes is also taken into account. The comparison between the IR-improved power spectrum \eqref{1loop1}, and the original 1-loop result \eqref{1loop2} can be seen in fig.~\ref{fig:power}. The IR-resummation clearly reduces the residual wiggles in the EFT prediction and can thus increase the range over which the theory agrees with simulations, as was pointed out in \cite{Senatore:2014via}.

\begin{figure}[!t]
\centering
\includegraphics[width=0.48 \textwidth]{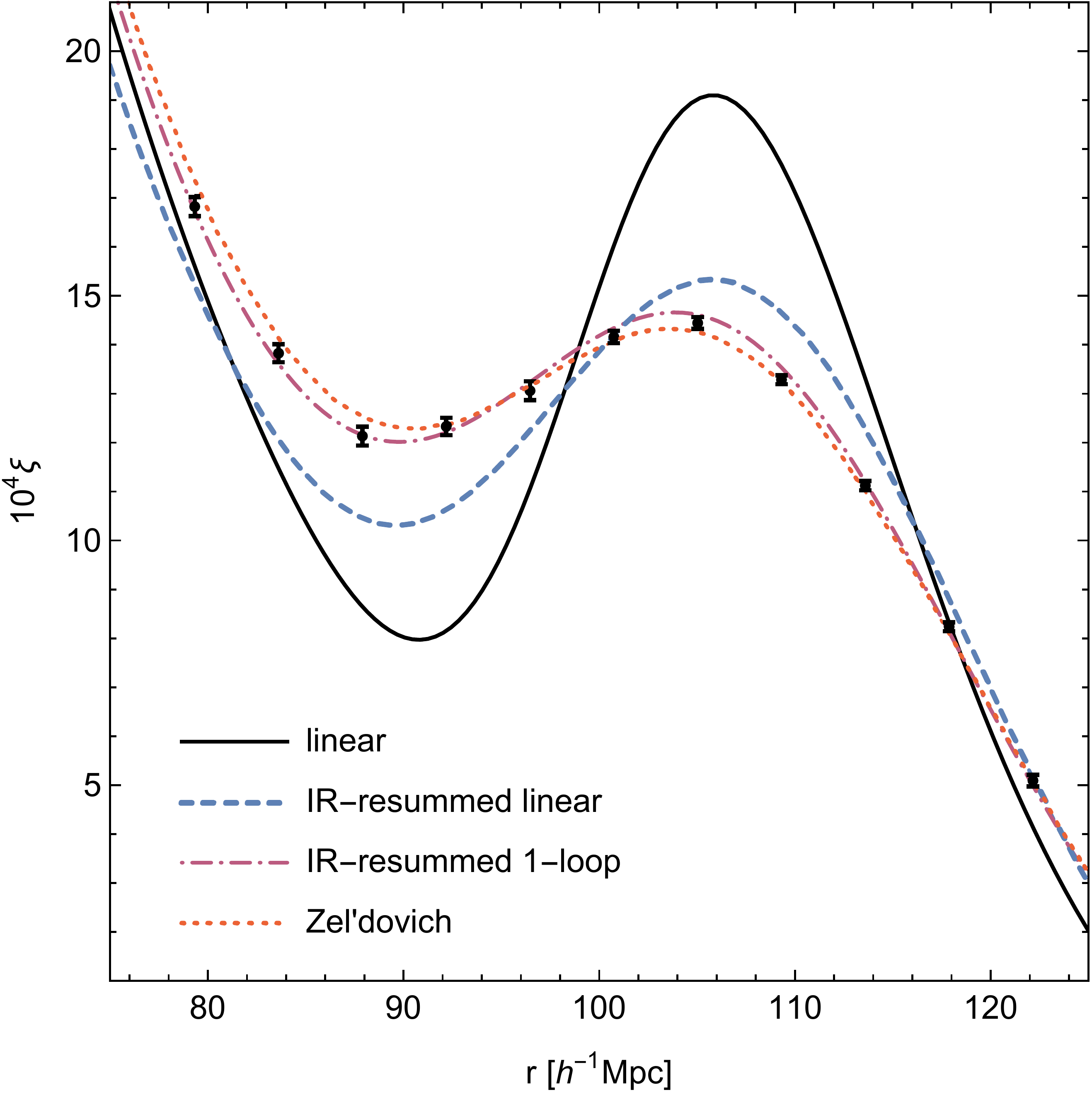}
\caption{\small {\em Various theoretical approximations to the acoustic peak in the correlation function as well as simulation measurements. Solid: linear, dashed: IR-resummed linear, dot-dashed: IR-resummed 1-loop, and dotted: Zel'dovich.}}
\label{fig:spread}
\end{figure}

For the correlation function, the broadened acoustic peak resulting from the IR-resummed linear and 1-loop power spectra is shown together with the initial peak in fig.~\ref{fig:spread}. Although the first approximation does not fully capture the smoothing of the peak seen in the data, it shows that indeed most of the spread is caused by the bulk motions. 

Without resummation the 1-loop EFT (or SPT) power spectra result in a spurious double-peaked feature at the BAO scale similar to the one shown in fig.~\ref{1Lspread}. This is because they only include $\Sigma_{\epsilon k}^2 \xi''(r)$ while higher derivative terms $1/n! \Sigma_{\epsilon k}^{2n}\xi^{(2n)}(r)$ that partially cancel this feature are not absent. The presence of this feature is the cause for the common wisdom that SPT does not work for the correlation function. As the good performance of the IR-resummed EFT proves, the failure is not related to the high-$k$ behavior of the perturbation theory but to the missing non-perturbative treatment of motions. One can indeed see that the IR-resummed EFT provides a good description of the correlation function down to $10\;h^{-1}{\rm Mpc}$ separations \cite{Senatore:2014via}. 

Another feature of fig.~\ref{fig:spread} that is worth emphasizing is the shift of the peak compared to the linear correlation function. This shift is expected to be due to corrections to $\tilde \xi_g$ of order $\Sigma^2 \xi_g'/\lb$, which are smaller than the broadening effects by a factor of $\sigma/\lb$ \cite{Sherwin}. They are not entirely fixed by symmetries since the cross correlation between a displacement and other nonuniversal effects --- e.g. arising from living in an over dense region --- caused by a long wavelength mode contributes at the same level. Nevertheless, they can be calculated in perturbation theory and are included, to leading order, in the 1-loop result, which predicts the position of the peak reasonably well. On the other hand, the BAO reconstruction schemes, to be discussed below, reproduce the original peak by virtue of undoing the displacements caused by the long modes which also eliminates the above mentioned cross correlations.

For comparison, we have also plotted in fig.~\ref{fig:spread} the Zel'dovich correlation function, which is known to give a relatively accurate description of the BAO spread. We will next argue that the success of the Zel'dovich approximation is because it can be formulated as \eqref{dressed}.

\vspace{0.3cm}
\noindent
{\bf Zel'dovich approximation.---}  The matter correlation function can be related to the correlation function of the relative displacement $\Delta \s(\z)$ of two points with initial (Lagrangian) separation $\z$:
\be\label{xilgr}
1+\xi(\x)=\int \frac{{\rm d}^3\kb}{(2\pi)^3} e^{i\kb\cdot \x}\int {\rm d}^3\z e^{-i\kb\cdot\z}\expect{e^{-\kb\cdot\Delta\s(\z)}}.
\ee
In the Zel'dovich approximation, $\Delta\s$ is replaced by its linear expression, and the above expectation value is trivially expressed in terms of the variance
\be\label{A}
\begin{split}
A^{ij}(\z)&=\expect{\Delta s^i(\z)\Delta s^j(\z)}\\[8pt]
&=\int\!\! {\rm d}^3\q \:\frac{q^i q^j}{q^4}P_{\rm lin}(q)\sin^2\left(\frac{\q\cdot\z}{2}\right).
\end{split}
\ee
Let us define Zel'dovich power spectrum as the result of the inner integral in \eqref{xilgr} at $k\neq 0$:
\be\label{Pz}
P_z(\kb)=\int {\rm d}^3\z e^{-i\kb\cdot\z}e^{-\frac12 A^{ij}(\z)k^i k^j},
\ee
which in the presence of the BAO feature contains an oscillating component $P_z^w(k)$. This can be approximated by the product of a non-smoothed piece times a broadening factor, as in \eqref{dressed}: Define $A_S^{ij}(\z,\Lambda)$, and $A_L^{ij}(\z,\Lambda)$ by the same integral as in \eqref{A}, but taken, respectively, over short modes $q>\Lambda$, and long modes $q<\Lambda$. So we have
\be
A^{ij}(\z)=A^{ij}_S(\z,\Lambda)+A^{ij}_L(\z,\Lambda).
\ee
A Zel'dovich power spectrum in the absence of the long modes $P_{z,S}(\kb,\Lambda)$, where $\Lambda\ll k$, can now be defined by replacing $A^{ij}\to A^{ij}_S$ in \eqref{Pz}. This is the analog of the last factor in \eqref{dressed}: it contains the full nonlinear effect of the short modes in the Zel'dovich approximation, but no long modes whatsoever.

Consider now the full $P_z(\kb)$. The integral in \eqref{Pz} is dominated by $\z=\Or(1/k)$, and, if $k$ is in the support of $P_z^w(k)$, by $\z=\pm\lb\hat\kb+\Or(1/k)$. The second contribution is what we called $P_z^w(\kb)$. Here, $A_L^{ij}(\z)$ is first of all appreciable, and second, it can be approximated to be a constant given by its value at $\z = \lb \hat\kb$ to yield
\be
\begin{split}
P_z^w(\kb)&\approx e^{-\frac 12 A^{ij}_L(\lb\hat\kb,\Lambda)k^i k^j} P_{z,S}^w(\kb,\Lambda)\\[8pt]
&\approx e^{-\Sigma^2_\Lambda k^2} P_{z,S}^w(\kb,\Lambda).
\end{split}
\ee
The second equality holds up to terms suppressed by $\sigma/\lb$. Replacing $\Lambda\to \ep k$ results in the desired analog of \eqref{dressed}. 

Hence, the Zel'dovich approximation, despite being a crude model of short scale dynamics, gives an accurate description of BAO broadening by taking into account the leading displacement caused by all longer wavelength modes on any given scale $k$.\footnote{Two alternative approximations have been proposed in the literature (e.g. \cite{Eisenstein:2006nj,Crocce:2007dt}) to model the broadening effects: 
\be\label{siginf}
P^w(\kb)\approx e^{-\Sigma^2_\infty k^2} P_{\rm lin}^w(k),
\ee
and
\be\label{sigv}
P^w(\kb)\approx e^{-\sigma_v^2 k^2} P_{\rm lin}^w(k),
\ee
where the velocity dispersion $\sigma_v^2$ is given by the same integral as in \eqref{Sigma} with $\Lambda = \infty$, but without the last square brackets. The two expressions happen to give similar results for the matter correlation function, and to be in good agreement with the result of simulations. However, we think the agreement in our Universe is accidental. The velocity dispersion is missing the factor $\sin^2(\q\cdot \x/2)$ in the relative displacement, which suppresses the contribution of the super-long modes. Had there been more power at large scales, or if $k_{\rm eq}\lb \ll 1$, \eqref{sigv} and \eqref{siginf} would have differed significantly. On the other hand, equation \eqref{siginf} approximates the short-long effects by the same expression as that of the long-short effects. This is not justified by any symmetry argument, and is an overestimation in the real universe. \eqref{siginf} would predict too much spreading if there was more power in small scales.}

\vspace{0.3cm}
\noindent
{\bf BAO reconstruction.---} This naturally leads us to the discussion of BAO reconstruction, and its connection to the long-short correlations \eqref{eq:real} and \eqref{bi}. The BAO reconstruction is a method to reproduce a sharper acoustic peak by undoing the bulk motion induced by the long wavelength modes, and hence, it is based on the same underlying idea that led to our results \cite{Eisenstein:2006nk,Padmanabhan:2008dd}. Given that the leading effect of the long mode is a uniform acceleration, this procedure roles back part of the time-evolution, which as we saw leads to the broadening of the BAO peak. 

Operationally, the reconstruction method consists of three steps: (i) Choosing a rigid separation $\Lambda$ between long and short modes, and solving for the linear displacement field produced by the long modes. (ii) Moving back all points according to this linear displacement field (as one would do in the Zel'dovich approximation). (iii) Adding back the original smooth field that is largely erased by step (ii). In this procedure, the only effect of the long modes on the short modes that has been reliably taken into account is the linear displacement. Hence, the effectiveness of the method seems to be a strong indication of the validity of \eqref{eq:real}. But, there are two caveats. First, the reconstruction method does not significantly affect the smooth part of the correlation function, hence it only verifies \eqref{eq:real} after background subtraction. 

\begin{figure}[!t]
\begin{center}
\includegraphics[width=0.48 \textwidth]{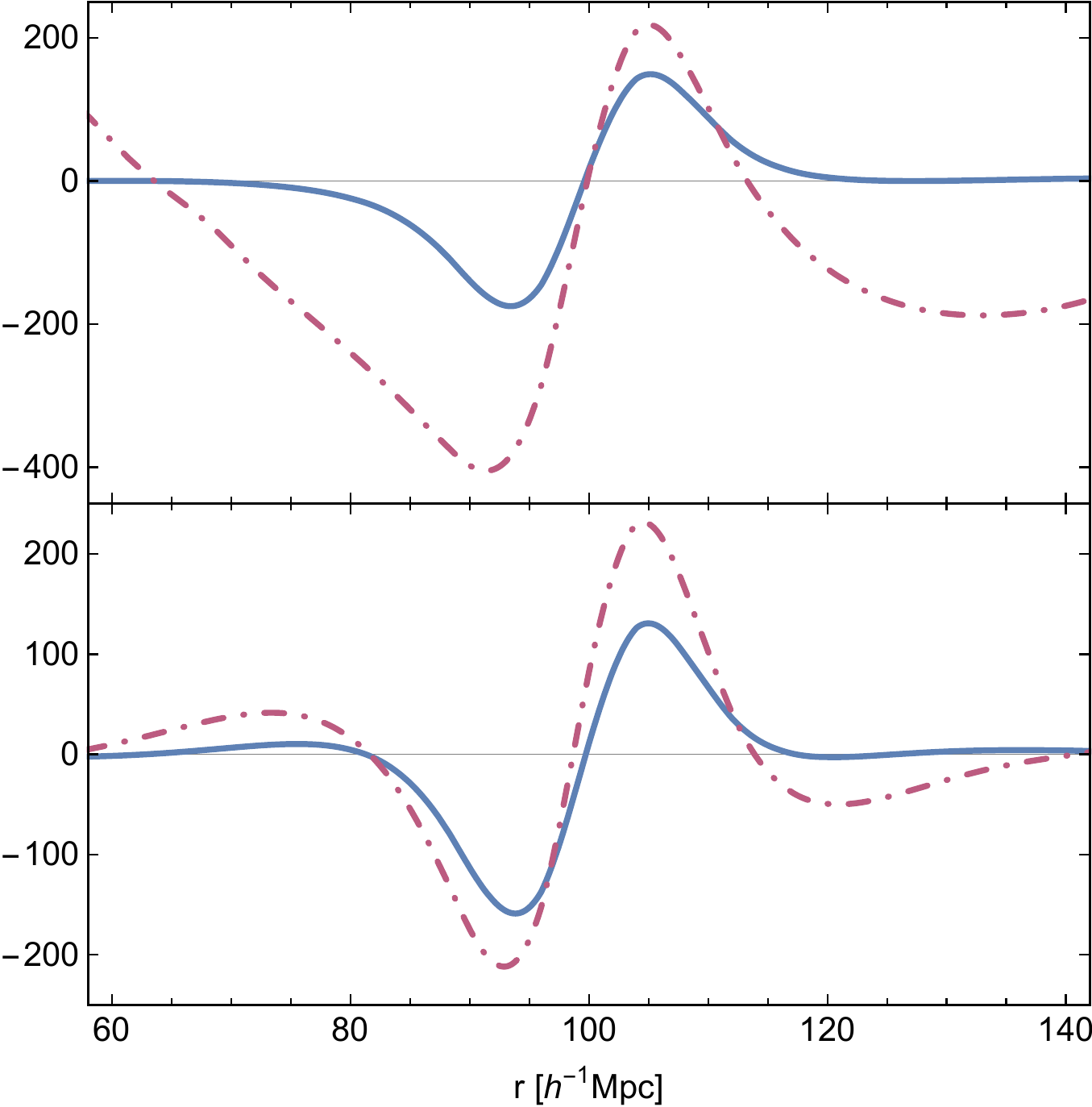}
\end{center}
\caption{\small {\em The same as fig.~\ref{fig:mixed} with $q=0.1\;h{\rm Mpc}^{-1}$. All curves are normalized by $P_{\rm lin}(k_{\rm eq})\xi(2\pi/k_{\rm eq})$. Upper panel: Without high-pass filter. Lower panel: With high-pass filter. In both cases the smooth background is subtracted.}}
\label{fig:mixedhp}
\end{figure}

Second, the threshold $\Lambda$ is practically chosen within the support of $P_g^w$, where as mentioned above, the modes both contribute to the peak, and cause it to spread. It is natural to suspect equation \eqref{eq:real} to become a poor approximation for these $q$ modes, due to their dynamical self-coupling. On the other hand, the reconstruction method would primarily deal with the effect of the $q$ modes on higher $k$ modes. Therefore, the effectiveness of reconstruction implies that even for these relatively larger values of $q$, once the contribution of modes below the threshold $\Lambda$ is removed from $\delta_g$, equation \eqref{eq:real} should be a good approximation.

To test this expectation in perturbation theory, we insert a high-pass filtered power spectrum $P_g(k)=(1-W(k,q))P_{\rm lin}(k)$ into the tree-level matter bispectrum, while keeping $P_{\rm lin}(q)$ unfiltered. The inverse Fourier transform with respect to $\kb$ is then to be compared to the r.h.s. of \eqref{eq:real}, with $\xi_g$ obtained from the same high-pass filtered $P_g(k)$. The results are shown in fig.~\ref{fig:mixedhp}, and seem to be in moderate agreement. The high-pass filter effectively picks small laboratories, free-falling in the background of the long wavelength mode. 

\section{Conclusions}

We used the leading Newtonian effect of a long wavelength matter perturbation $\delta_L$ to derive approximate formulas for its correlation with the distribution of pairs \eqref{eq:real}, as well as the squeezed limit bispectrum \eqref{bi}, in the presence of the BAO feature. The derivation is based only on two underlying assumptions: first, the equivalence principle, by which we imply that no additional (fifth) force is universally sourced by material objects, and second, local formation of tracers which forbids nontrivial bias with respect to locally unobservable quantities such as velocity and gravitational potential. This requires absence, or rather smallness, of primordial local non-Gaussianity. Therefore, the result holds beyond the standard perturbation theory, and apply equally well to biased tracers. In the real Universe, it gives the dominant component of the real space correlation at $x\sim \lb$, but a subdominant--though still unique and distinguishable--piece in momentum space. 

Next, we explored the connection with the broadening of the acoustic peak, where the same universal effect but averaged over the long modes is known to account for most of the spread in the observed Universe. We derived a formula for the observed correlation function \eqref{dressed}, which resums the induced motion by the long modes to all orders. A simpler way to implement this IR-resummation in perturbation theory was proposed, and the result was shown to be in good agreement with the numerical results from a $N$-body simulation, and with the Zel'dovich approximation. It was shown that the Zel'dovich approximation to the correlation function can be recast into the form of our IR-resummed formula \eqref{dressed}, which we take as the explanation for its success in predicting the BAO spread. Finally, we discussed BAO reconstruction method as a practical application of the same underlying idea. 


\vspace{0.3cm}
\noindent
{\em Acknowledgements.---} We thank P. Creminelli for useful discussions. T.B. acknowledges support from the Institute for Advanced Study through the Corning Glass Works Foundation Fund. M.M. is supported by NSF Grants PHY-1314311 and PHY-0855425. M.S. acknowledges support from the Institute for Advanced Study. M.Z. is supported in part by the NSF grants  PHY-1213563 and AST-1409709.

\end{document}